\documentclass[a4paper]{jpconf}
\usepackage{graphicx}
\usepackage{xcolor}
\begin{document}
\title{Multiplicity fluctuations of identified hadrons in p+p interactions at SPS energies.}
\vspace*{-0.3cm}
\author{Maja Ma\'{c}kowiak-Paw{\l}owska$^{1,2}$ and Andrzej Wilczek$^3$\\ for the NA61 Collaboration}
\address{$^1$Goethe University, Max-von-Lause-Str. 1, 60438, Frankfurt am Main, Germany}
\address{$^2$Warsaw University of Technology, ul. Koszykowa 75, 00662 Warsaw, Poland}
\address{$^3$University of Silesia, ul. Bankowa 12, 40007 Katowice, Poland}
\vspace*{-0.05cm}
\ead{majam@if.pw.edu.pl, awilczek@us.edu.pl}
\vspace*{-0.3cm}
\begin{abstract}
Study of energy and system size fluctuations of identified hadrons is
one of the key goals of NA61/SHINE at the CERN SPS. Results may allow
to discover the critical point (CP) of strongly interacting matter as
well as to uncover properties of the onset of deconfinement (OD).
But fluctuations exhibit numerous other sources starting
from most basic ones like volume effects and conservation laws.

NA49 seems to observe fluctuations related to CP in collisions of
medium size nuclei at top SPS energy. However, this result will remain
inconclusive until systematic data on energy and system size dependence
will be available. Moreover, fluctuations in p+p as well as in Pb+Pb interactions
should be better understood.

In this contribution results on multiplicity fluctuations of identified hadrons in
p+p interactions at the CERN SPS energies will be presented. The NA61
data will be compared with the corresponding results from central
Pb+Pb collisions of NA49 in the common acceptance region of both
experiments. Moreover, predictions of models (EPOS, UrQMD and HSD) for
both reactions will be tested.
\end{abstract}
\vspace{-0.5in}
\section{Introduction}
It is a well established fact that matter exists in different states. One of the most important goals of heavy-ion collision (HIC) experiments is to study the phase diagram of strongly interacting matter by finding possible phase boundaries and establishing their properties. In particular, high energy HIC experiments want to study the deconfined matter (QGP) and the transition between QGP and confined hadrons. The NA61/SHINE experiment~\cite{na61} is a fixed target experiment in the European Organization for Nuclear Research (CERN) at the Super Proton Synchrotron (SPS). The study of energy and system size dependence of fluctuations of identified hadrons is one of the~goals of the NA61/SHINE ion program. The aim of this program is to search for the critical point (CP)~\cite{CP} of strongly interacting matter as well as to uncover properties of the onset of deconfinement (OD)~\cite{SMES,OD}. The predecessor of NA61/SHINE, the NA49 experiment~\cite{na49} seems to observe fluctuations possibly related to the CP in collisions of medium size nuclei at the top energy of SPS~\cite{QM_Kasia}. However, measured fluctuations are affected by numerous effects not related to CP or OD like volume fluctuations and conservation laws. Thus, this result will remain inconclusive until systematic data on energy and system size dependence will be available. These data are being recorded and analysed by NA61/SHINE.
\vspace*{-0.35cm}
\section{NA61/SHINE identified hadron fluctuation studies}
\begin{figure}[h]
\begin{center}
\begin{minipage}{14pc}
\includegraphics[width=14pc]{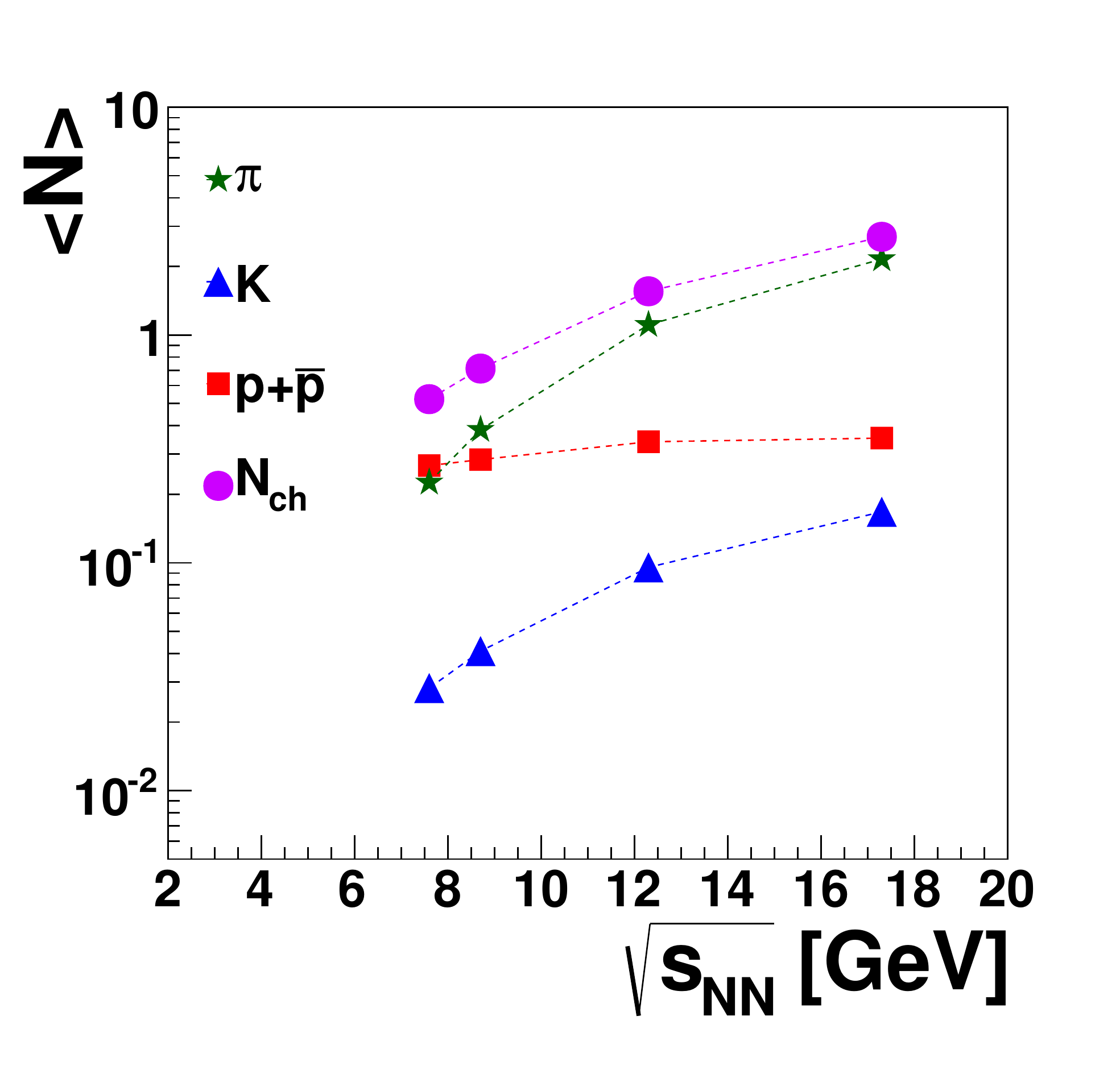}
\vspace{-0.3in}
\end{minipage}\hspace{2pc}%
\begin{minipage}{14pc}
\includegraphics[width=14pc]{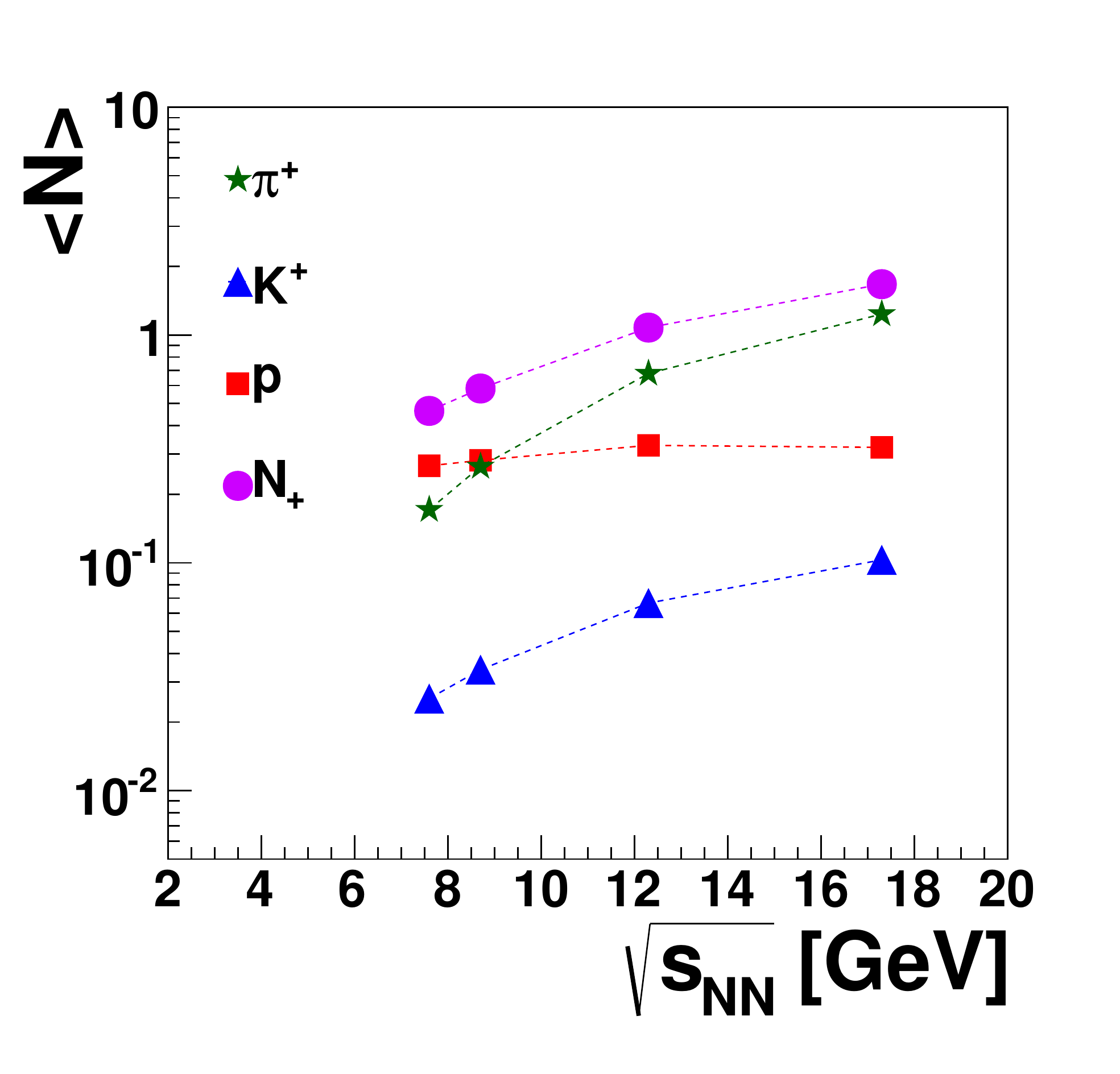}
\vspace{-0.3in}
\end{minipage}
\end{center} 
\vspace*{-0.25cm}
\caption{\label{n_all} The energy dependence of mean multiplicity of identified charged (\textit{left}) and positively charged (\textit{right}) hadrons in p+p interactions in the acceptance used for the fluctuation analysis.}
\vspace*{-0.3cm}
\end{figure}

Addressed, multiplicity and chemical fluctuations of identified hadrons produced in strong and electromagnetic processes were measured in inelastic p+p interactions at $\sqrt{s_{NN}}=\ 7.6$, 8.7, 12.3, and 17.3 GeV. Particle identification is based on energy loss measurements in the relativistic rise region. Fluctuations of charged pions ($\pi=\pi^{+}+\pi^{-}$), kaons ($K=K^{+}+K^{-}$) and protons + antiprotons as well as of positively charged hadrons were studied via first and second (pure and mixed) moments of identified particle multiplicity
distributions. For the completeness, figure~\ref{n_all} shows energy dependence of mean multiplicity of identified all charged and positively charged hadrons in the analysis acceptance. Note that the mean multiplicities are not corrected for possible biases and should not be used for direct comparison with models. 

Second moments of identified hadron multiplicity distributions were obtained using the identity method~\cite{identity1,identity2, identity3}. Presented results include the statistical uncertainty and a first estimate of systematic uncertainty (still under studies are detector effects and influence of feed-down).\newline
\begin{figure}[h]
\vspace*{-0.2cm}
\begin{center}
\begin{minipage}{14pc}
\includegraphics[width=14pc]{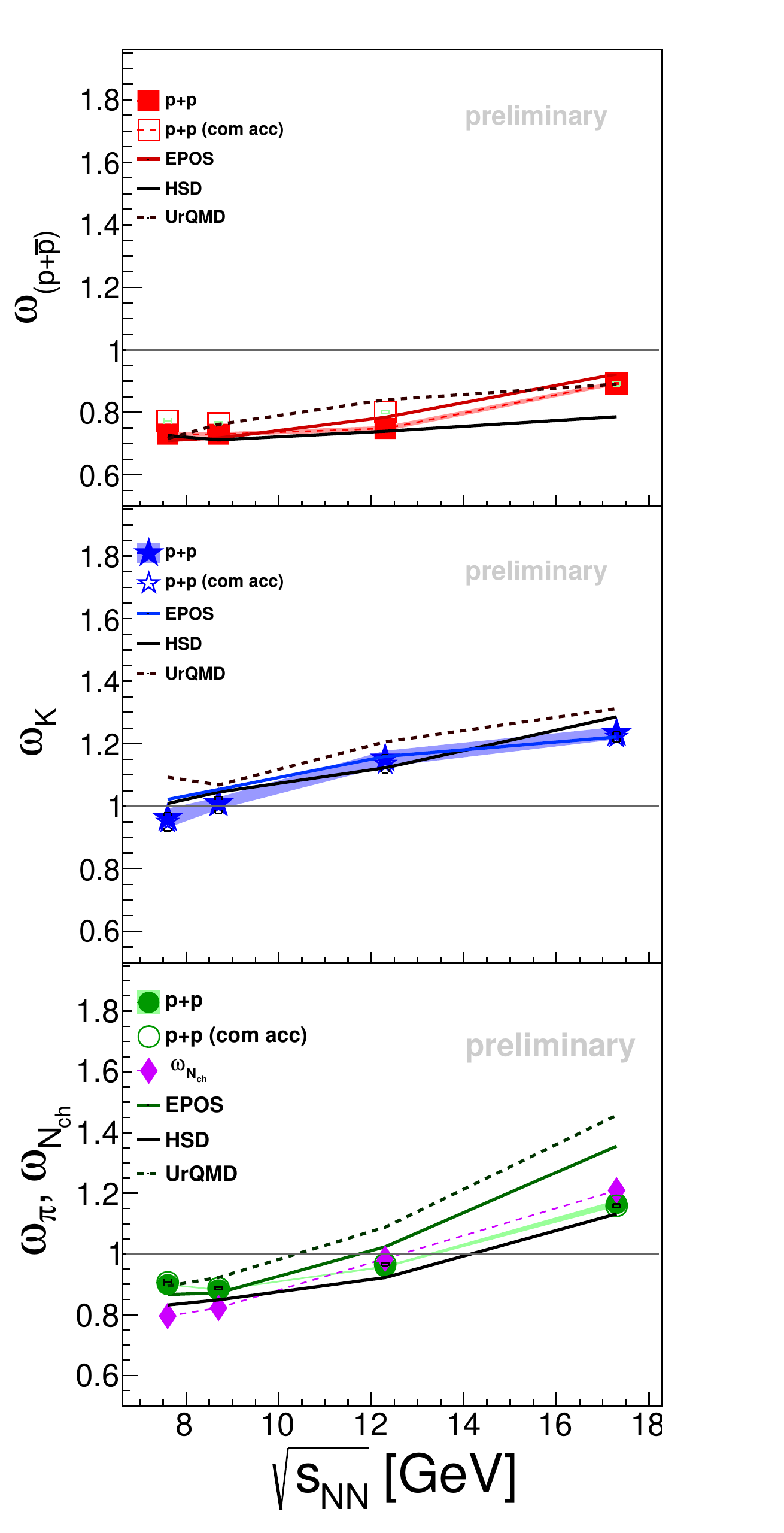}
\vspace{-0.3in}

\end{minipage}\hspace{2pc}%
\begin{minipage}{14pc}
\includegraphics[width=14pc]{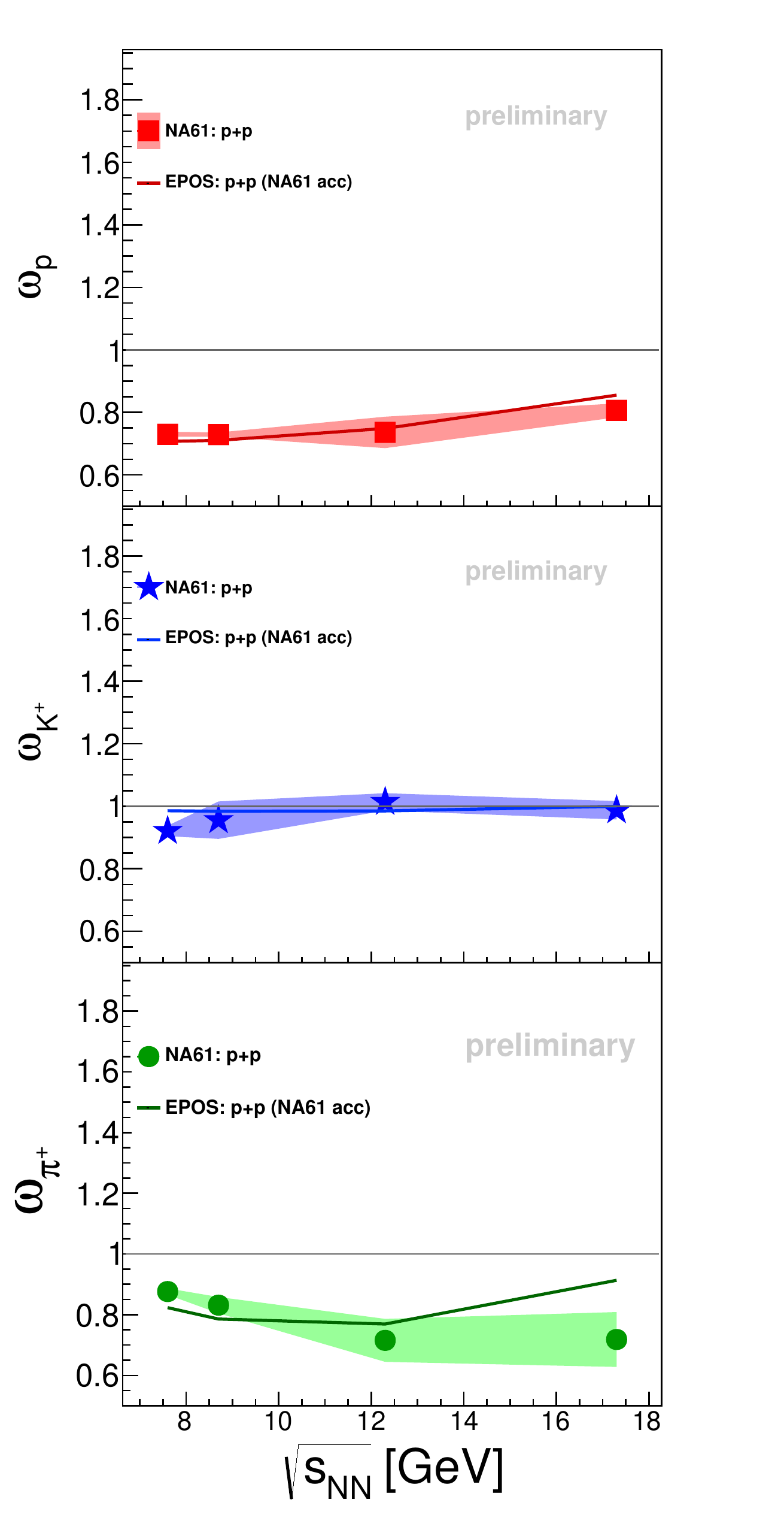}
\vspace{-0.3in}
\end{minipage} 
\end{center}
\vspace*{-0.2cm}
\caption{\label{omega_all} The energy dependence of the scaled variance in p+p interactions: $\omega_{p+\bar{p}}$, $\omega_{K}$, $\omega_{\pi}$~(\textit{left}), $\omega_{p}$, $\omega_{K^{+}}$, and $\omega_{\pi^{+}}$~(\textit{right}).
Full symbols denote the fluctuation results within the NA61 acceptance. Open symbols denote the results within the common phase space region of NA61 and NA49. Model predictions are presented for the NA61/SHINE acceptance.}
\vspace*{-0.6cm}
\end{figure}
The multiplicity fluctuations were addressed using the scaled variance measure, $\omega_{i}$ of the multiplicity distribution. It is defined as $\omega=\frac{\langle N_{i}^{2}\rangle-\langle N_{i}\rangle^{2}}{\langle N_{i}\rangle}$, where $\langle N_{i}\rangle$ and $\langle N_{i}^{2}\rangle$ are the mean multiplicity and the second moment of the multiplicity distribution of particles of type $i$, respectively. The scaled variance is an intensive measure~\cite{SIQ}, i.e. it is independent of the number of wounded nucleons in the Wounded Nucleon Model~\cite{WNM} or volume
in the Grand Canonical Ensemble, but it depends on their fluctuations. The latter feature makes it difficult to compare results from p+p interactions with those from nucleus+nucleus collisions. For the Poisson distribution $\omega=1$. \newline
Figure~\ref{omega_all} (\textit{left}) shows the scaled variance of $\pi$, $K$ and $p+\bar{p}$. The scaled variance of all considered particle types increases with increasing energy. Rich data on the charged particle multiplicity distribution in the full phase space obey the KNO scaling~\cite{KNO,KNO2}. From this scaling follows a linear increase of the scaled variance with the mean multiplicity of charged particles visible as an increase of $\omega_{\pi}$ with collision energy (see the bottom left panel of Fig.~\ref{omega_all}). The scaled variance of unidentified charged hadrons measured within the acceptance chosen for this analysis is shown in the same panel by purple diamonds. The increase with collision energy is weaker than that measured in the full phase space, but agreement between $\omega_{N_{ch}}$ and $\omega_{\pi}$ as well as between $N$ and $N_{\pi}$ indicates that KNO scaling may also apply to pion production.\newline
Figure~\ref{omega_all} (\textit{right}) shows the energy dependence of $\omega_{p}$, $\omega_{K^{+}}$, and $\omega_{\pi^{+}}$. The scaled variance of identified positively charged hadrons is smaller than the one measured for the sum of charges. Moreover, in the case of positively charged hadrons, instead of an increase with the collision energy the scaled variance is approximately independent of energy.\newline
The suppressed ($<$1) values of $\omega_{p+\bar{p}}$ and $\omega_{p}$ are probably due to the baryon number conservation~\cite{CE}. In p+p interactions, the net baryon number is two, thus, fluctuations of the sum of charges is dominated by protons~\cite{CE}. Values of $\omega_{K}$ equal to or above 1 are probably connected with the strangeness conservation, which leads to a correlation between the $K^{+}$ and $K^{-}$ productions. It is supported by the fact that $\omega_{K^{+}}<\omega_{K}$~\cite{kaony}. It should be underlined that the influence of a specific charge conservation on the scaled variance depends on the volume of the system, mean number and mass of charge carriers, and the fraction of measured particles and anti-particles. Measured quantities are approximately reproduced by the EPOS~\cite{EPOS,EPOS2}, UrQMD~\cite{UrQMD,UrQMD2} and HSD~\cite{HSD2,HSD3,HSD4} models.
\begin{figure}[h]
\begin{center}
\begin{minipage}{14pc}
\begin{center}
\includegraphics[width=14pc]{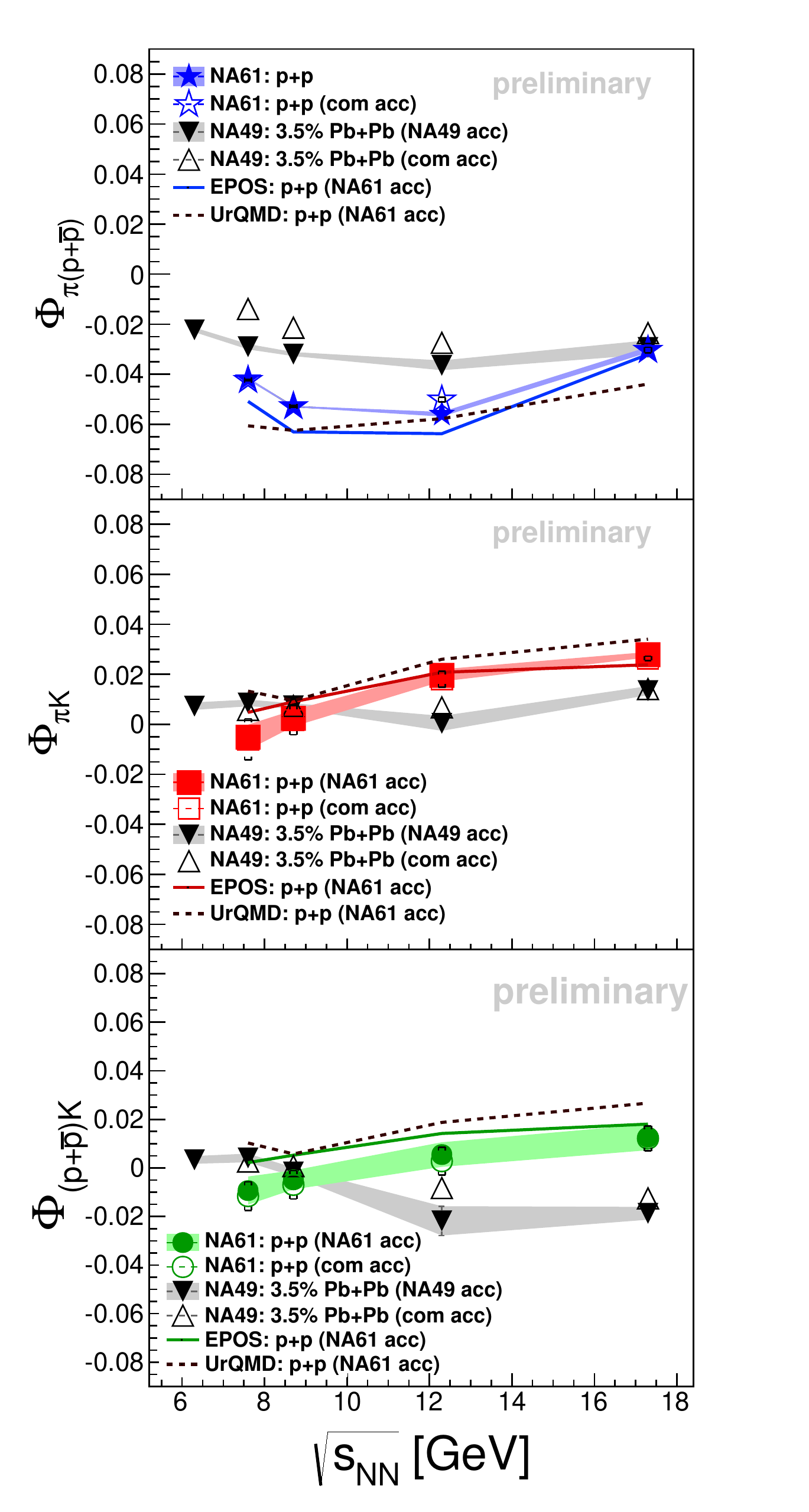}
\vspace{-0.3in}
\end{center}
\end{minipage}\hspace{2pc}%
\begin{minipage}{14pc}
\begin{center}
\includegraphics[width=14pc]{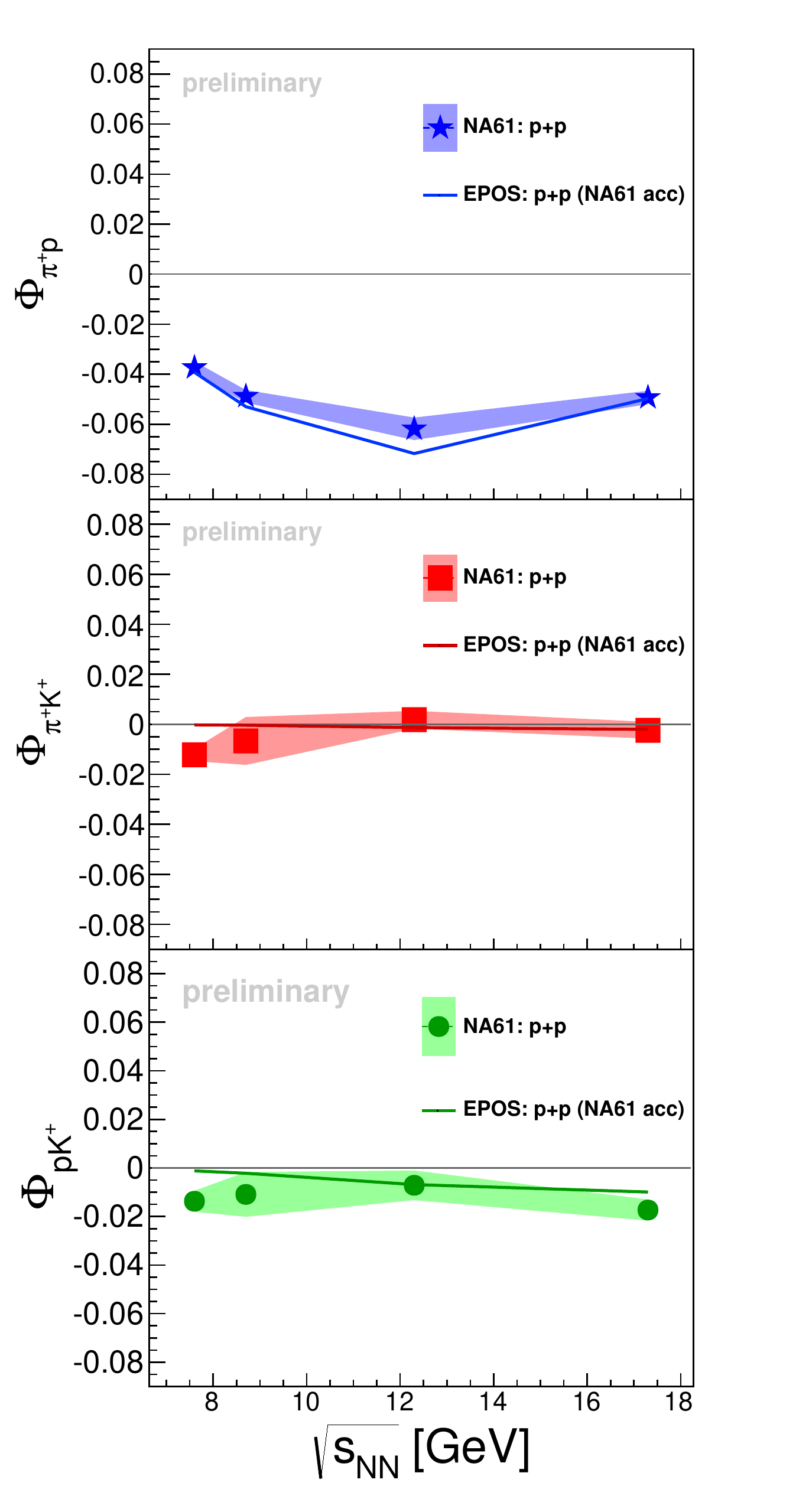}
\vspace{-0.3in}
\end{center}
\end{minipage} 
\end{center}
\vspace*{-0.3cm}
\caption{\label{phi_all} The energy dependence of $\Phi_{\pi(p+\bar{p})}$, $\Phi_{\pi K}$, $\Phi_{(p+\bar{p})K}$~(\textit{left}), $\Phi_{\pi^+ p}$, $\Phi_{\pi^{+} K^{+}}$, and $\Phi_{pK^{+}}$ (\textit{right}) in p+p interactions.
Full symbols denote fluctuation results within the NA61 or NA49 acceptance. Open symbols denote results within the common phase space region of NA61 and NA49. Model predictions are presented for the NA61/SHINE acceptance.}
\vspace*{-0.6cm}
\end{figure}
In order to compare results for p+p and $3.5\%$ of most central Pb+Pb collisions~, the~strongly intensive measure, 
$\Phi_{ij}=\frac{\sqrt{<N_i><N_j>}}{<N_i+N_j>}\times\left(\sqrt{\Sigma^{ij}}-1\right)$~\cite{Phi,SIQ}, defined for two hadron types, $i$ and $j$, was chosen, where $\Sigma^{ij}=[<N_i>\omega_j+<N_j>\omega_i-2(<N_iN_j>-<N_i><N_j>)]/<N_i+N_j>$.\\
As a strongly intensive measure, $\Phi_{ij}$ is not only independent of number of wounded nucleons or volume but also of their fluctuations. For independent particle emission, $\Phi_{ij}=0$. Figure~\ref{phi_all} shows the energy dependence of $\Phi_{ij}$ of two hadron types: $\pi(p+\bar{p})$, $\pi K$, and $(p+\bar{p})K$ as well as only for positively charged particles.  Full symbols refer to the NA61 respectively NA49 acceptance. Open symbols display results for the common phase-space region of NA49 and NA61. Differences between the individual and common acceptance are small. For $\pi(p+\bar{p})$ there is a dip at intermediate energies. The energy dependence is similar in the case of positively charged hadrons. A similar but weaker effect is visible in Pb+Pb interactions. In the case of $\pi K$ and $(p+\bar{p})K$ there is an increase of $\Phi_{ij}$ with collision energy, which is not visible in positively charged hadrons. The increase of $\Phi_{\pi K}$ is not visible in Pb+Pb collision but $\Phi_{\pi K}\geq0$ at all considered energies in both reactions. The opposite energy dependencies between p+p and Pb+Pb interactions are visible in $\Phi(p+\bar{p})K$ (see bottom panel of Fig.~\ref{phi_all}~(\textit{left})).

\end{document}